\documentclass[aps,pre,reprint,amsmath,amssymb,superscriptaddress]{revtex4-1}
\usepackage{morefloats}
\usepackage{bm}
\newcommand{\beq}{\begin{equation}}
\newcommand{\eeq}{\end{equation}}

\usepackage{comment}

\usepackage[retainorgcmds]{IEEEtrantools}
\usepackage{graphicx,tikz,placeins}
\usepackage{mathrsfs}
\usepackage{wasysym}
\usepackage{amsmath,amssymb,amsfonts,physics}
\usepackage{color}

\usepackage{float}

\usepackage{times,txfonts}
\usepackage{nicefrac}

\usepackage{lipsum}
\usepackage[colorlinks=true,linkcolor=blue,urlcolor=blue,citecolor=blue,pdfusetitle]{hyperref}
\usepackage{physics}
\usepackage{soul}

\usepackage{ulem}
\newcommand{\mom}[1]{\langle#1\rangle}

\begin{document}

\title{Collective heat engines via different interactions: Minimal models, thermodynamics and phase transitions}

\author{Iago N. Mamede, Vit\'oria T. Henkes and Carlos E. Fiore }
\affiliation{Universidade de São Paulo,
Instituto de Física,
Rua do Matão, 1371, 05508-090
São Paulo, SP, Brazil}

\date{\today}
\begin{abstract}We investigate the dynamics and thermodynamics
of a framework 
composed of interacting units in which 
parameters (temperatures and energies)  
assume distinct values
due to the contact with distinct (cold and hot) thermal reservoirs. 
The influence
of different  ingredients, such as the contact between thermal baths
(simultaneous versus not simultaneous contact), the coupling between them
(equal or different couplings) and the topology of interactions (all-to-all and local interactions) is investigated.
Closed expressions for transition lines have been obtained, expressed by a linear combination of interaction energies times reciprocal  temperatures for the simultaneous thermal contact baths and deviates from it when the contact is not simultaneous. The interplay between performance and dissipation is investigated under different conditions,  giving rise to a richness of operation regimes, such as heat-engine and heat pump. The relationship  between thermodynamic quantities  (power, efficiency and dissipation) allows a careful choice of parameters to ensure the desirable compromise between them.
Finally, the influence of  different interactions energies (Ising, Potts versus Blume-Emery-Griffiths (BEG) like) are investigated, revealing
that Potts interactions in general present superior performances than BEG ones.
\end{abstract}

\maketitle
\section{Introduction}
The study of  thermal engines is the cornerstone of equilibrium thermodynamics~\cite{carnot1978reflexions,callen} and the advent of stochastic thermodynamics has extended to nanoscopic systems~\cite{seifert2012stochastic}
such concept under firmer basis, giving rise to new analytical tools and methodologies~\cite{crooks,fiorek,karel2016prl,harunariPhysRevX.12.041026,PhysRevE.110.064126,barato2015thermodynamic,RevModPhys.91.045001,deffner2020thermodynamic,pancotti2020speed,2206.02337} as well as different optimization strategies~\cite{verley2014unlikely, schmiedl2007efficiency, cleuren2015universality, van2005thermodynamic, esposito2010quantum, seifert2011efficiency, izumida2012efficiency, golubeva2012efficiency, holubec2014exactly, bauer2016optimal, karel2016prl, tu2008efficiency, ciliberto2017experiments, bonanca2019, mamede2021obtaining, karel2016prl}.
While  a plenty of aforementioned nanoscopic systems are  composed of a single unit, the extension to  few \cite{akasaki2020entropy,mamede2021obtaining} or  several interacting systems, such as  complex networks~\cite{bonifazi2009gabaergic,schneidman2006weak,buzsaki2014log,gal2017rich,gatien,herpich,Herpich_2020,mamede2023,forao2023powerful,forao2024splitting}, biological structures~\cite{rapoport1970sodium, gnesotto2018broken,lynn2021broken,smith2019public},  quantum systems~\cite{mukherjee2021many,niedenzu2018cooperative,kurizki2015quantum,lee2022quantum,campisi2016power,mukherjee2020universal,halpern2019quantum,kim2022photonic,PhysRevApplied.19.034023,PhysRevResearch.1.033192,chen2019interaction} has attracted a great deal of interest, not only for the possibility of boosting the system performance but also for the
presence of novel features, such as different sort of nonequilibrium phase transitions  ~\cite{hooyberghs2013efficiency,campisi2016power,mamede2023}. In this context, Ising \cite{yeomans1992statistical,PhysRevE.82.031104} and 
Potts \cite{Potts1951,Potts1952,RevModPhys.54.235} models are highlighted by
their simplicities for capturing the fundamental aspects of ferromagnetism.
Different nonequilibrium versions  of them have been proposed and investigated, giving rise to  a richness of  phenomena, such as
phase transitions \cite{martynec2020entropy,fiore2021current,felipe1,tome2023} and different critical behaviors \cite{forao2024splitting}, optimized work-to-work converters \cite{herpich,herpich2} and heat-engines, both able to present
superior perfomances  \cite{gatien,forao2023powerful,forao2024splitting} than common single units
setups \cite{filho21,forao2025characterization}.

In this contribution, we  investigate the dynamics and the thermodynamic properties (power, efficiency and dissipation) of a class of collective
systems placed in contact with two thermal reservoirs, whose parameters (e.g. individual and interaction energies)  assume distinct values
depending on the bath contact. In all cases, the interplay between parameters (temperatures  and interaction energies) leads to different operation regimes, such as heat engine and pump, whose scaling behaviors of  thermodynamic quantities (power, efficiency and dissipation) are characterized by maximum power and a simple form for efficiency, allowing a careful choice of parameters ensuring the desirable compromise for the performance. We analyze in details the influence of different ingredients, such as  the coupling between thermal baths (equal or different couplings), the topology of interactions (all-to-all and local interactions) and the non-simultaneous contact between thermal baths. While the simultaneous
contact characterizes phase transitions via a bilinear relation of model parameters times reciprocal temperatures (hence uncovering simple different routes for reaching the phase transition), transition points acquire a more intricate dependence on model parameters when the contact is not simultaneous. Both descriptions become equivalent for fast switching between thermal baths.

The outline of this paper is structured as follows: in Sec.~\ref{Model} we present the model, while the thermodynamics of simultaneous contact is described in Sec.~\ref{simult}
for equal and different couplings, as well as for all-to-all and interactions forming regular arrangements. The extension for non-simultaneous coupling between thermal reservoirs is considered in Sec.~\ref{nonsimult} and conclusions are drawn in Sec.~\ref{Conclusions}.

\section{Model}\label{Model} 
Let us consider a generic system composed of  an ensemble of $N$ interacting units, in which each unit $i\in \{1,...,N\}$
has spin $s_i$  belonging to the set $\mathcal{S}$ of all possible values.  
For example, $s_i\in\mathcal{S}=\{0,1,...,q-1\}$ for the $q-$state Potts model
and $s_i\in\mathcal{S}=\{0,\pm\}$ for the spin-1 BEG \cite{beg}.
System energies can be decomposed in the following form
\begin{equation}
    E^{(\nu)}(\bm{s})=E^{(\nu)}_{\rm ss}(\bm{s})+E^{(\nu)}_{\rm ds}(\bm{s})+E^{(\nu)}_{\rm ind}(\bm{s}),
    \label{General_Energy}
\end{equation}
where  $\bm{s}=\{s_1,s_2,...,s_N\}$ and
$E^{(\nu)}_{\rm ind}(\bm{s})$, $E^{(\nu)}_{\rm ss}(\bm{s})$ and $E^{(\nu)}_{\rm ds}(\bm{s})$ denote  the contribution of individual units and the interaction terms between units in same and different states, respectively. Each one is associated with  the contact with $\nu$-th
thermal bath and given by
\begin{align}
    E^{(\nu)}_{\rm ind}(\bm{s})&=\sum_{i=1}^N\sum_{\ell\in\mathcal{S}}\epsilon_\ell^{(\nu)}\delta_{s_i,\ell},\\
    E^{(\nu)}_{\rm ss}(\bm{s})&=-\frac{1}{2k}\sum_{(i,j)}\sum_{\ell\in\mathcal{S}}V_\ell^{(\nu)}\delta_{s_i,s_j},\\
    E^{(\nu)}_{\rm ds}(\bm{s})&=-\frac{1}{2k}\sum_{(i,j)}\sum_{\ell\in\mathcal{S}}\sum_{r\neq\ell}\varepsilon_{\ell,m}^{(\nu)}\delta_{s_i,\ell}\delta_{s_j,r},
\end{align}
where $k$  denotes the number of nearest neighbor of each unit $i$. 
Above expressions become simpler for the Potts and BEG models by
the fact that $\varepsilon_{\ell,m}^{(\nu)}=\epsilon_\ell^{(\nu)}=0$
(Potts), for any $\ell$ and $m$, and
 $\varepsilon_{+,-}^{(\nu)}=\varepsilon_{-,+}^{(\nu)}=-V^{(\nu)}_\ell$ (BEG) and $0$ otherwise. Thus, the interaction energies can be written as
\begin{align}
    E^{(\nu)}(\bm{s})&=-\frac{V_\nu}{2k}\sum_{(i,j)}\delta_{s_i,s_j},\label{energy1}\\
    E^{(\nu)}(\bm{s})&=-\frac{V_\nu}{2k}\sum_{(i,j)}s_is_j-h_\nu\sum_{\{s_i\}}s_i+\Delta_\nu\sum_{i=1}^Ns_i^2,
    \label{energy}
\end{align}
for Potts and BEG models, respectively, where $E^{(\nu)}(\bm{s})$
in the latter case was expressed in terms of new variables
 $\delta_{s_i,\pm}=(s_i^2\pm s_i)/2$ and $\delta_{s_i,0}=1-s_i^2$, where 
 $h_\nu=(\epsilon_+^{(\nu)}-\epsilon_{-}^{(\nu)})/2$ and 
$\Delta_\nu=-(\epsilon_+^{(\nu)}+\epsilon_{-}^{(\nu)})/2$.
It is immediate to see that the Ising model corresponds to the $q=2$ Potts model.
In all cases, we are dealing with one-site the dynamics, in which at each the time-evolution
of configuration $\bm{s}$ gives rise to $\bm{s'}\equiv (s_1,\dots,s_i',\dots,s_N)$, with $s_i\neq s_i'$, meaning that only the state of an individual unit can be changed at each time instant. 
Moreover, transition rates have the following Arrhenius form
\begin{equation}
    \omega^{(\nu)}_{\bm{s}',\bm{s}}=\Gamma_\nu \exp\left(-\frac{\beta_\nu}{2}\Delta\mathcal{E}^{(\nu)}_{\bm{s}',\bm{s}}\right),
    \label{Transition_Rate}
\end{equation}
with $\Gamma_\nu$ the connectivity between the system and the $\nu$-th reservoir, $\Delta\mathcal{E}^{(\nu)}_{\bm{s}',\bm{s}}\equiv E^{(\nu)}(\bm{s}')-E^{(\nu)}(\bm{s})$ the energy difference between configurations and $\beta_\nu=1/T_\nu$ the inverse temperature (for now on, we set $k_B=1$).

\section{Thermodynamics of two simultaneous thermal reservoirs}\label{simult}
For the simultaneous contact between thermal baths, the probability distribution ${p}_{\bm{s}}(t)$  at the time $t$ is governed by the following master equation
\begin{equation}
    d_t{p}_{\bm{s}}(t)=\sum_{\nu=1}^2\sum_{\bm{s}'\neq\bm{s}}J^{(\nu)}_{\bm{s'},\bm{s}}(t),
    \label{Master_Equation}
\end{equation}
where $J^{(\nu)}_{\bm{s'},\bm{s}}(t)$ denotes the probability current defined by
\begin{equation}
    J^{(\nu)}_{\bm{s'},\bm{s}}(t)=\omega^{(\nu)}_{\bm{s},\bm{s}'}p_{\bm{s}'}(t)-\omega^{(\nu)}_{\bm{s}',\bm{s}}p_{\bm{s}}(t),
\end{equation}
where $\omega^{(\nu)}_{\bm{s}',\bm{s}}$ is the transition rate from $\bm{s}$ to $\bm{s}'$ due to the contact with the $\nu$-th thermal bath. 
The dynamics evolves to a nonequilibrium steady-state in which $p_{\bm{s}}(t)\rightarrow p^{\rm st}_{\bm{s}}$ and
$J^{(\nu)}_{\bm{s}',\bm{s}}(t)\rightarrow \mathcal{J}^{(\nu)}_{\bm{s}',\bm{s}}$. 

In order to set up the system Thermodynamics, we consider  the entropy production expression \cite{schnakenberg1976network}
\begin{equation}
    \mom{\dot{\sigma}(t)}=\sum_\nu\sum_{\bm{s}'>\bm{s}}J_{\bm{s}',\bm{s}}^{(\nu)}(t)\log\left[\frac{\omega^{(\nu)}_{\bm{s'}\bm{s}}p_{\bm{s}}(t)}{\omega^{(\nu)}_{\bm{s}\bm{s}'}p_{\bm{s}'}(t)}\right],
    \label{eps_Gen}
\end{equation}
that acquires the following form in the NESS: $\mom{\dot{\sigma}}=\sum_{\nu=1}^2\mom{\dot{\sigma}_\nu}$, where
$\mom{\dot{\sigma}_\nu}$ is given by
\begin{align}
\mom{\dot{\sigma}_\nu}=&\sum_{\bm{s},\bm{s}'\neq\bm{s}}\mathcal{J}^{(\nu)}_{\bm{s}',\bm{s}}\log\frac{\omega^{(\nu)}_{\bm{s}',\bm{s}}}{\omega^{(\nu)}_{\bm{s},\bm{s}'}}.
    \label{SS_General_Res}
\end{align}
By inserting Eq.~(\ref{Transition_Rate}) into above expression together the previous property for steady fluxes, one arrives at  the following expression
\begin{equation}
\mom{\dot{\sigma}}=-\sum_\nu\beta_\nu\sum_{\bm{s},\bm{s}'\neq\bm{s}}\Delta\mathcal{E}^{(\nu)}_{\bm{s}',\bm{s}}\mathcal{J}^{(\nu)}_{\bm{s}',\bm{s}}.
    \label{SS_General}
\end{equation}
Above relation presents a "Clausius form"  for the steady
entropy production
$\mom{\dot{\sigma}}=-\sum_\nu\beta_\nu\mom{\dot{Q}_\nu}$. From Eq.~(\ref{SS_General})
together the first law of Thermodynamics
$\mom{\mathcal{P}}+\mom{\dot{Q}_1}+\mom{\dot{Q}_2}=0$, one arrives to the following expressions for the exchanged heat $\mom{\dot{Q}_\nu}$ and the power $\mom{\mathcal{P}}$
\begin{align}
\mom{\dot{Q}_\nu}&=\sum_{\bm{s},\bm{s}'\neq\bm{s}}\Delta\mathcal{E}^{(\nu)}_{\bm{s}',\bm{s}}\mathcal{J}^{(\nu)}_{\bm{s}',\bm{s}},\label{Heat_General}
\end{align}
and $\mom{\mathcal{P}}=-(\mom{\dot{Q}_1}+\mom{\dot{Q}_2})$.
We show, by means of different examples, that the simultaneous contact between thermal baths is signed by order-disorder phase transitions  as 
a control parameter $X_{\nu c}$ ($X \in \{\beta_\nu,V_\nu\}$ (Potts) and $X \in \{\beta_\nu,V_\nu,h_\nu,\Delta_\nu\}$ (BEG)  for $\nu=1$ or $2$) is varied.  In both cases,  transition points obey
 the generic bilinear relation
\begin{equation}
A_1\beta_1 V_{1}+A_2\beta_{2} V_{2}=\Psi,
\label{Phase_General1}
\end{equation}
where $A_1$ and $A_2$ and $\Psi$ depend on model parameters, $\beta_\nu$'s and $\Gamma_\nu$'s. It is immediate to see that Eq.~(\ref{Phase_General1})
reduces to the equilibrium phase transition $\beta^{\rm eq}V^{\rm eq}=\Psi/2$ as $\beta_1=\beta_2$, $X^{(1)}=X^{(2)}$ for all $X$
and $\Gamma_1=\Gamma_2$. In the next
sections, we shall derive $\Psi$ for Potts and BEG models for all-to-all
interactions.

\subsection{Equal couplings $\Gamma_1=\Gamma_2$}

\subsubsection{All-to-all dynamics and phase transitions}
A first insight about the role of each
ingredient shall be investigated  for all-to-all
interactions. It constitutes a simplified description of system
properties in which the dynamics becomes fully
characterized by the collection of number of units $N_i$ in each possible state 
(e.g. $\{N_0,N_1,...,N_{q-1}\}$ and $\{N_-,N_0,N_{+}\}$ for Potts and BEG models, respectively) 
where $\sum_{\alpha}N_\alpha=N$. By setting $k\rightarrow N$ and expressing Eqs.~(\ref{energy1})-(\ref{energy}) in terms of
 $N_\alpha$'s, one obtains the following expressions
\begin{align}
    E^{(\nu)}&=-\frac{V_\nu}{2N}\sum_{\alpha=0}^{q-1}N_\alpha(N_\alpha-1),\\
    E^{(\nu)}&=-\sum_{\alpha\in\{-,+\}}\left[\frac{V_\nu}{2N} N_\alpha(N_\alpha-1)-\Delta_\nu N_\alpha\right]\\
    +&\frac{V_\nu}{N}N_{+}N_{-}-h_\nu(N_+-N_-)\nonumber,
    \label{modelsN}
\end{align}
for Potts and BEG models, respectively, where, from now on, we shall set $h_\nu=0$ for all $\nu$.
For one-site dynamics, each transition $\alpha\rightarrow \alpha'$ corresponds  to $N_\alpha\rightarrow N_\alpha-1$ and $N_{\alpha'}\rightarrow N_{\alpha'}+1$, whose energy differences depend linearly of the difference of $N_\alpha$'s, as shown in Appendix~\ref{apa}. 
In the thermodynamic limit $N\rightarrow \infty$, the dynamics depends on densities \( {\bar n}_{\alpha} = \langle N_{\alpha}/N \rangle \) ($\alpha\in \{0,...,q-1\}$ and $\alpha\in \{0,\pm\}$
for Potts and BEG, respectively) \cite{gatien,forao2023powerful,herpich,herpich2,mamede2023} and governed by the following master equation
\begin{equation} 
\dot{{\bar n}}_\alpha(t) = \sum_{\nu=1}^2 \sum_{\alpha' \neq \alpha} \{ \omega^{(\nu)}_{\alpha \alpha'} {\bar n}_{\alpha'}(t) - \omega^{(\nu)}_{\alpha' \alpha} {\bar n}_\alpha(t)\}, 
\label{mee}
\end{equation}  
 where although akin to Eq.\eqref{Master_Equation},  Eq.~(\ref{mee}) 
 acquires a non-linear form. 

The all-to-all interactions, the steady entropy production $\mom{\dot{\sigma}}$
is then given by
 \begin{equation}
\mom{\dot{\sigma}}=\sum_{\nu}\sum_{\alpha'>\alpha}\{ \omega^{(\nu)}_{\alpha \alpha'} {\bar n}_{\alpha'} - \omega^{(\nu)}_{\alpha' \alpha} {\bar n}_\alpha\}\log\left[\frac{\omega^{(\nu)}_{\alpha'\alpha}}{\omega^{(\nu)}_{\alpha\alpha'}}\right],
    \label{eps_NESS}
\end{equation}
also presenting the Clausius like form 
$\mom{\dot{\sigma}}=-\sum_\nu\beta_\nu\mom{\dot{Q}_\nu}$.
\textit{Potts model—}
A reliable order-parameter $m$ in this case is related to
densities $\overline{n}_\alpha$ through relation  \cite{RevModPhys.54.235}
\begin{equation}
\overline{n}_0=\frac{1}{q}[1+(q-1)m],~~~ \overline{n}_{\alpha\neq 0}=\frac{(1-m)}{q}.
\end{equation}
From Eq.~(\ref{mee}) for $i=0$, together expressions above, the time evolution
of $\overline{n}_0(t)$ is given by
\begin{equation}
\dot{\overline{n}}_0(t)=\left[\sum_{\alpha>0}^{q-1}(\omega_{0\alpha}^{(1)}+\omega_{0\alpha}^{(2)})\overline{n}_\alpha(t)\right]-\left[\sum_{\alpha>0}^{q-1}(\omega_{\alpha0}^{(1)}+\omega_{\alpha0}^{(2)})\right]\overline{n}_0(t),
\label{ppp}
\end{equation}
whose nonequilibrium steady-state regime $m(t)\rightarrow m$ is described by a simple expression and given by
\begin{equation}
    m=\frac{2}{2-q+q \coth \left[\frac{1}{4} \left(\beta _1V_1 +\beta _2V_2
   \right)m\right]}.
   \label{PottsMinimum}
\end{equation}

\begin{figure*}[htb!]
    \centering
    \includegraphics[width=1\linewidth]{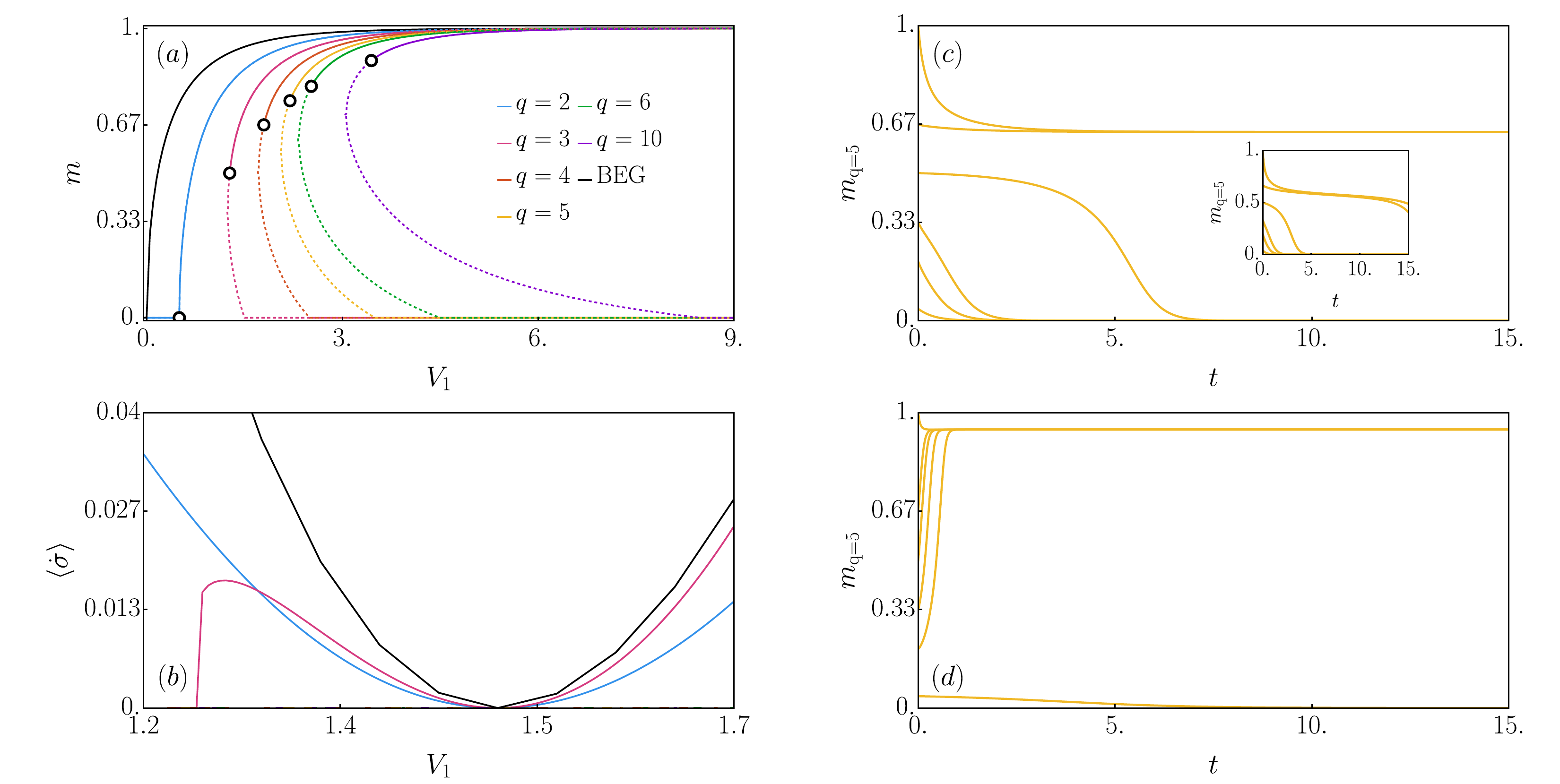}
    \caption{Order parameter $m$ (a) and the entropy production $\langle {\dot \sigma}\rangle$ (b) versus $V_1$ for the all-to-all q-state Potts and the BEG models ($\Delta_1=\Delta_2=0$), respectively. Dots correspond to the transition point
    from Eq.~\eqref{qGT2BetaV} (Potts) and $\beta_1V_1+\beta_2V_2=3$ (BEG). Panels (c) and (d) show, for $q=5$, the time evolution of for $m(t)$ different initial conditions for  $V_{1c}=2.089...$ [from Eq.~\eqref{qGT2BetaV}] and $V_1\approx V_{1f}$, respectively. Inset: The same but for $V_1=2.06<V_{1b}$ Parameters: $V_2=3$, $\Gamma_\nu=1$, $\beta_1=2$ and $\beta_2=1$.}
    \label{Figure_Phase_Transition}
\end{figure*}

From Eq.~(\ref{eps_NESS}), the entropy production $\mom{\dot{\sigma}}$ is given by
\begin{equation}
\label{epa}
    \mom{\dot{\sigma}}=\frac{ (q-1) m\left(\beta _2 V_2-\beta _1 V_1\right) \sinh \left[\frac{1}{4} m \left(\beta _2 V_2-\beta _1 V_1\right)\right]}{q e^{\frac{1}{4} m \left(\beta _1
   V_1+\beta _2 V_2\right)}-2 \sinh \left[\frac{1}{4} m \left(\beta _1 V_1+\beta _2 V_2\right)\right]},
\end{equation}
irrespectively the value of $q$. The behavior of $  \mom{\dot{\sigma}}$
is exemplified in Fig.~\ref{Figure_Phase_Transition}c.

As mentioned previously,  the
system undergoes a phase transition for 
all values of $q$ when a control parameter $X \in\{\beta_1,\beta_2,V_1,V_2\}$ is tuned, being continuous for $q=2$ and discontinous
for $q\ge 3$. In the former case, $m$ acquires the simple form $m=\tanh(\frac{1}{4} \left(\beta _1 V_1 +\beta _2V_2 \right)m)$, showing a critical point yielding at $\beta _1V_1 +\beta _2V_2=2q$ and consistent
with Eq.~(\ref{Phase_General1}) for $\Psi=2q$ and $A_1=A_2=1$.
Although Eq.~(\ref{PottsMinimum})
depicts the jump of $m$
and the existence of a spinodal region for $q\ge 3$---trademarks of discontinuous
phase transitions--- it hides the precise location of phase coexistence point because there is no free-energy  to decide which phase the system is.
In order to  characterize the phase transition for $q\ge 3$, we analyze three indicators of discontinuous transition points:
the values $V_{1b}$  and $ V_{1f}$ which delimit the
bistable region (the system evolves to the steady solution $m(t)\rightarrow 0$
and $m(t)\rightarrow m_0\neq 0$ as $t\rightarrow \infty$, irrespective the initial condition) and also by means of expression
\begin{align}
    \beta_1V_1+\beta_2V_2=\frac{4(q-1)}{q-2}\log(q-1).\label{qGT2BetaV}
\end{align}
It can be understood by taking
the equilibrium case
as $\beta_1=\beta_2$ and $V_1=V_2$, whose phase coexistence point
yields at $\beta V=\Psi/2$ \cite{RevModPhys.54.235}.
 Fig.~\ref{Figure_Phase_Transition} exemplifies above
features for distinct $q$'s and fixed values of $\beta_1,\beta_2$ and $V_2$.
Transition points (symbols $\circ$) clearly follows   Eq.~(\ref{Phase_General1}) for $q=2$.
Athough hysteretic branches also follow Eq.~(\ref{Phase_General1})  for $q\ge 3$,
in such a case $\Psi \neq 4(q-1)\log(q-1)/(q-2)$).
 Finally,  the entropy production  jumps
from  $\mom{\dot{\sigma}}_c$ to $0$, where  $\mom{\dot{\sigma}}_c$ solely depend
on $q$ and parameters.

\textit{BEG model—}
The analysis of  the BEG model is more revealing due
to the presence of individual energies $\Delta_\nu \neq 0$'s favoring
states $\pm$ with respect to $0$.  Its dynamics can be described via order parameter $m=\overline{n}_{+}-\overline{n}_{-}$ and  $\rho=\overline{n}_{+}+\overline{n}_{-}$. Again for the $\Gamma_1=\Gamma_2$ case, from Eq.~(\ref{mee}), the steady-state solution $( m,\rho)$ is obtained by solving the non-linear system of equation
\begin{equation}
    \mathcal{K}_1 m =  \log\left[\frac{\rho + m}{\rho - m}\right],\quad
    \mathcal{K}_2 = \log\left[\frac{4(1 - \rho)^2}{\rho^2 - m^2}\right],
    \label{BEG_minimum}
\end{equation}
where $\mathcal{K}_1\equiv\beta_1V_1+\beta_2V_2$ and $\mathcal{K}_2\equiv\beta_1\Delta_1+\beta_2\Delta_2$.
In order to characterize the criticality, it is convenient to write $\rho=\rho(m)$ and by expanding the left side of Eq.~(\ref{BEG_minimum})
into power series, from which one arrives at steady state expression $0=\phi_1m+\phi_3m^3+\phi_5m^5+\dots$, where  
 coefficients read
\begin{equation}
\phi_1=\beta_\nu \frac{a_1}{a_2}(X_\nu-X_{\nu c}),\qquad \phi_3=\frac{a_1a_2}{24}\left[4-e^{\frac{1}{2} \mathcal{K}_2}\right],
\label{tric}
\end{equation}
with $a_1=4+e^{\frac{\beta_1 \Delta_1}{2}}+e^{\frac{\beta_2\Delta_2}{2}}$,  $a_2=2+e^{\frac{1}{2}(\beta_1\Delta_1+\beta_2\Delta_2)}$ and the critical
point
$X_{\nu c}$ is  given by $\beta_1V_1+\beta_2V_2=2+e^{\frac{1}{2}(\beta_1\Delta_1+\beta_2\Delta_2)}$, again satisfying Eq.~(\ref{Phase_General1}) where $\Psi=2+e^{\frac{1}{2}(\beta_1\Delta_1+\beta_2\Delta_2)}$ 
and $A_1=A_2=1$.


\textbf{}

\subsubsection{Power,  efficiency and dissipation}\label{Power_etc}

\begin{figure}
    \centering
    \includegraphics[width=.9\linewidth]{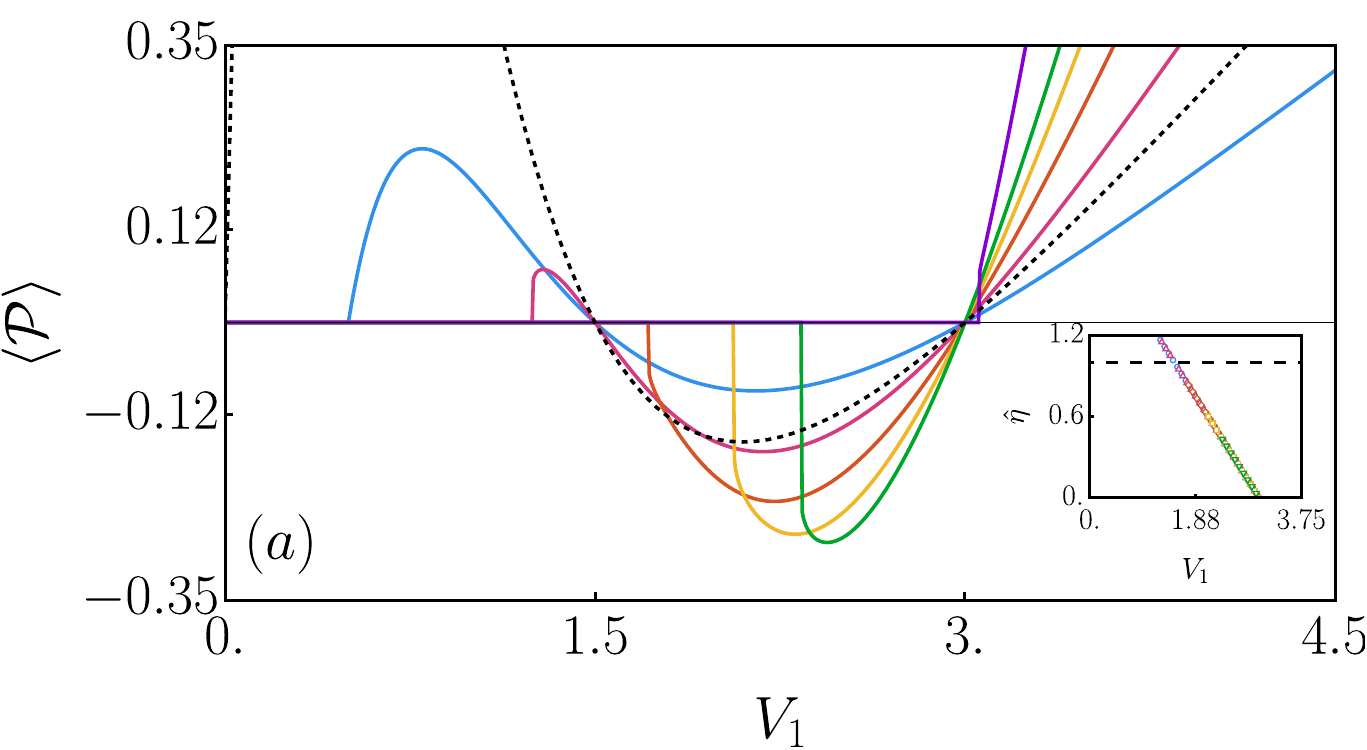}
    \includegraphics[width=.9\linewidth]{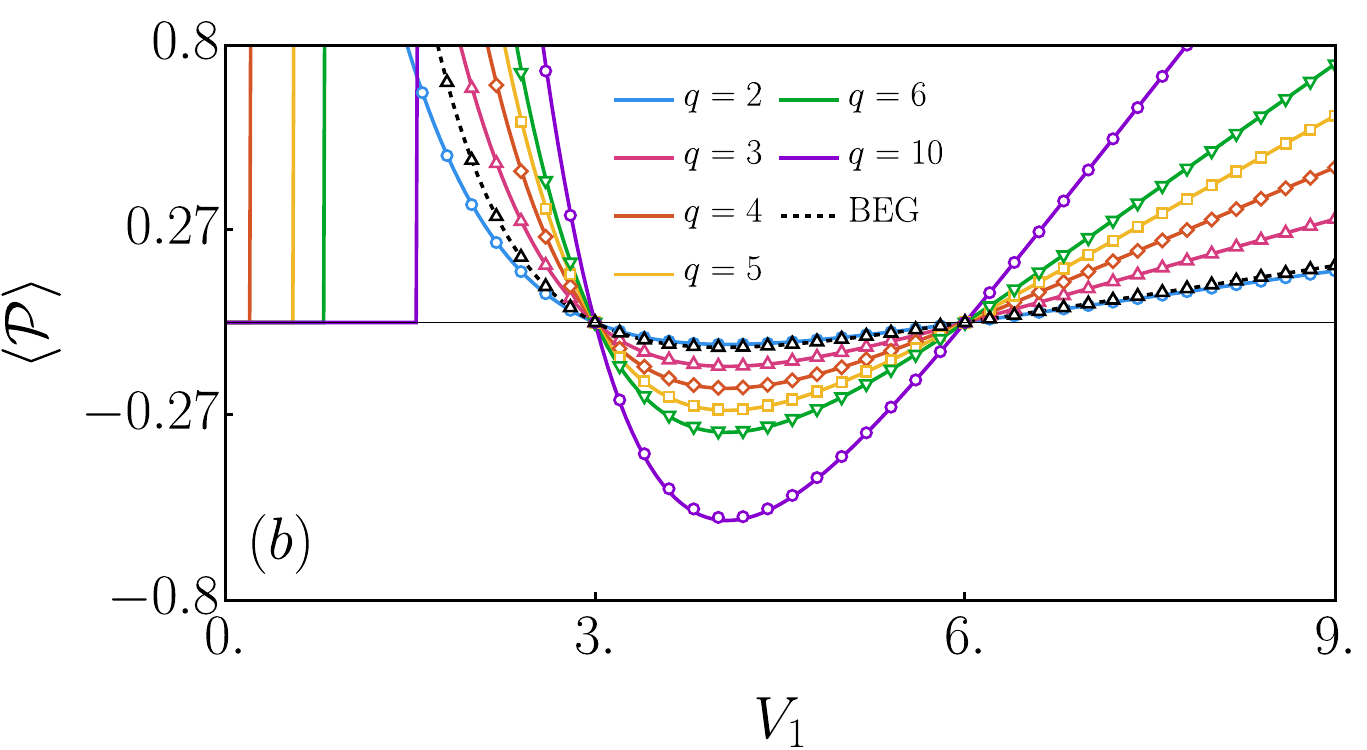}
    \caption{Depiction of the power $\mom{\mathcal{P}}$ versus $V_1$ for different values of $q$ (Potts) and for $\Delta_1=\Delta_2=0$ (BEG),  for $V_2=3$ (a) and $V_2=6$ (b). In the panel (b), the lines and symbols correspond to the exact solution and phenomenological description, respectively. Inset: Normalized efficiency $\hat{\eta}=\eta/\eta_c$ versus $V_1$ for the same models and parameters of the main panel (a). Parameters: $\Gamma_\nu=1$, $\beta_1=2$ and $\beta_2=1$.}
    \label{Power_Grid}
\end{figure}

Above class of systems can exhibit different operation regimes (heat-engine, heat-pump and dud) as parameters are properly tuned and depending on which phase the system is. To see this, we start our analysis for the all-to-all Potts model,
in which expressions 
for the exchanged heat $\langle {\dot Q}_\nu\rangle$ and power $\langle {\cal P}\rangle$ are somewhat simpler and given by
\begin{equation}
    \langle {\dot Q}_\nu\rangle=\frac{ (q-1)  (-1)^{\nu+1}V_\nu m \sinh \left[\frac{1}{4} m \left(\beta _2 V_2-\beta _1 V_1\right)\right]}{q e^{\frac{1}{4} m \left(\beta _1
   V_1+\beta _2 V_2\right)}-2 \sinh \left[\frac{1}{4} m \left(\beta _1 V_1+\beta _2 V_2\right)\right]},
   \label{heat2}
\end{equation}
and 
\begin{equation}
\label{power}
 \langle {\cal P}\rangle  =\frac{ (q-1) (V_2-V_1) m \sinh \left[\frac{1}{4} m \left(\beta _2 V_2-\beta _1 V_1\right)\right]}{q e^{\frac{1}{4} m \left(\beta _1
   V_1+\beta _2 V_2\right)}-2 \sinh \left[\frac{1}{4} m \left(\beta _1 V_1+\beta _2 V_2\right)\right]},
\end{equation}
respectively. We pause to make a few comments: First, the efficiency, given by $\eta=- \langle {\cal P}\rangle/\langle {\dot Q}_2\rangle$, acquires the simple form
\begin{equation}
\eta=1-\frac{V_1}{V_2},
\label{eff}
\end{equation}
which is independent on temperatures and $q$.
 Second,  from Eq.~(\ref{power}) we see that the roots of power yield at $V_1=V_2$  and $V_1/V_2=\beta_2/\beta_1$, consistent with
a heat-engine in  which $\langle {\cal P}\rangle<0$ and $\langle {\dot Q}_2\rangle>0$  as  $V_1/V_2>\beta_2/\beta_1$, irrespectively the value of $q$.
Third, the former ($V_1=V_2$) and latter ($V_1/V_2=\beta_2/\beta_1$) roots bound the heat-engine regime,
corresponding to the minimum of power fluctuations and dissipation,
respectively, as given by Eq.~(\ref{epa}) and stated  recently in Ref.~\cite{forao2025characterization}.
Fourth, the heat-pump ($\langle {\cal P}\rangle>0$ and $\langle {\dot Q}_2\rangle<0$ ) operation   yields for $V_1/V_2<\beta_2/\beta_1$. Fifth and last,  the system operates at ideal efficiency $\eta_c$ as  $V_1/V_2=\beta_2/\beta_1$, provided     $\beta_1V_1+\beta_2V_2>\Psi$. 
Since $\langle {\cal P}\rangle= \langle {\dot Q}_1\rangle= \langle {\dot Q}_2\rangle=0$ for $\beta_1V_1+\beta_2V_2\le \Psi$, the phase transition will imply to
discontinuous behaviors of thermodynamic quantities in the case
of the phase transition occurring between above roots of $\langle {\cal P} \rangle$. Fig.~\ref{Power_Grid} illustrates all such above features for different $q$'s and representative values of interactions ($V_2=3$ and $V_2=6$) for $\beta_2=2$, $\beta_2=1$ (analogous findings
 for $V_\nu$'s held fixed and  $\beta_\nu$ be varied). As discussed previously, the heat-engine regimes are constrained between $V_1=V_2$ [$V_1=3(6)$] in the left~(right) panels and $V_1=\beta_2V_2/\beta_1$ [$V_1=3$ in the bottom panel], being
 independent on $q$ and exhibit $-\langle {\cal P}\rangle$'s increasing as $q$ is varied, above all
its maximum value $-\langle {\cal P}\rangle_{mP}$'s. Taking into account that top panels depict phase transitions yielding
at $V_{1c}=\left(3-4(q-1)\ln(q-1)/(q-2)\right)/2<\beta_2V_2/\beta_1=3/2$ [evaluated according to Eq.~(\ref{qGT2BetaV})], the phase transition shortens
the heat-engine operation for $q\ge 3$, which is dependent on $q$ in such a case.
Similar results are obtained by estimating the transition point from $V_{1b}$.

\begin{figure}
    \centering
    \includegraphics[width=1.\linewidth]{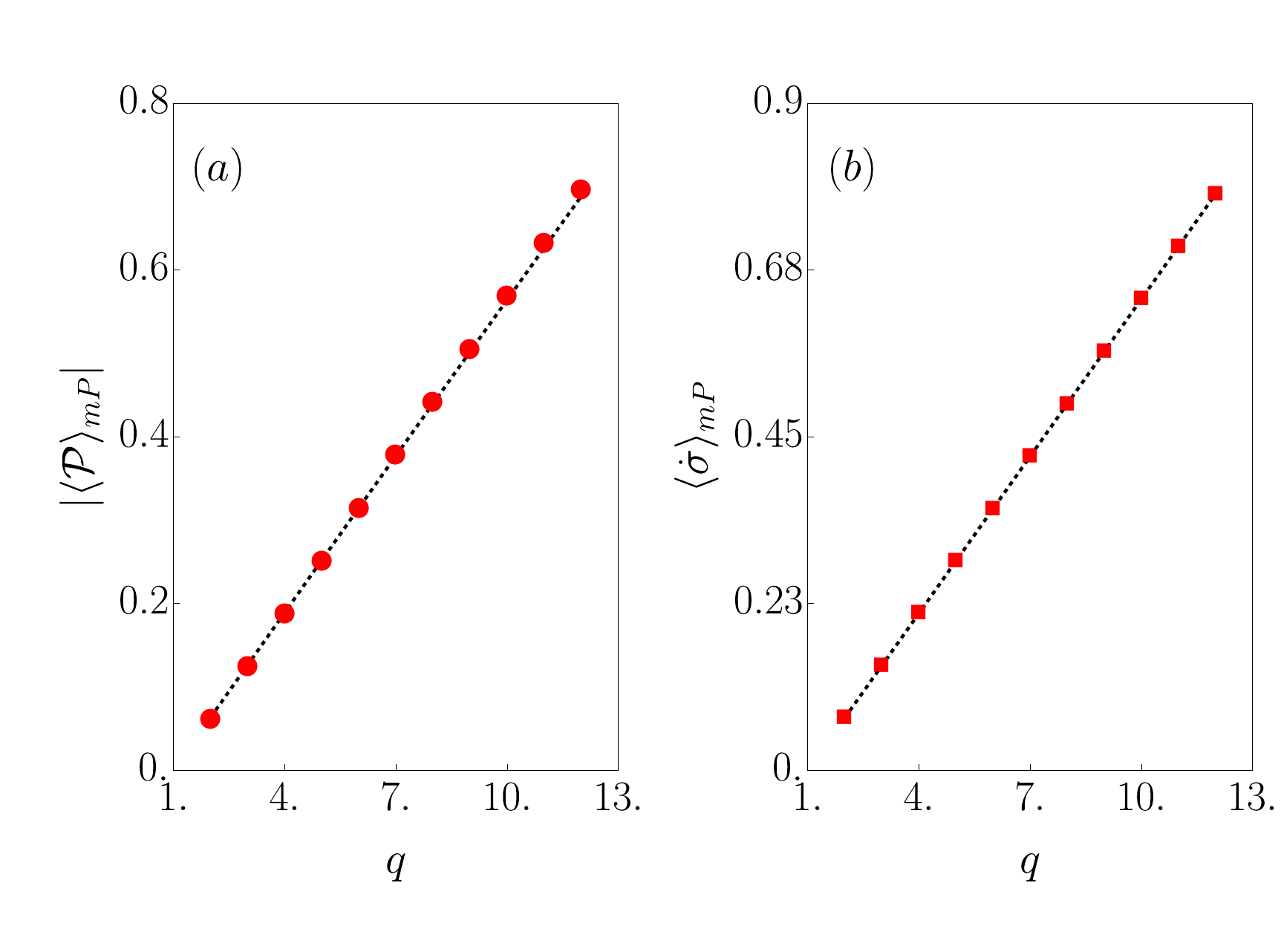}
    \caption{Panels (a) and (b) depict the maximum power 
    $-\mom{\mathcal{P}}_{mP}$ (with respect to $V_1$)   and corresponding entropy production  $\mom{\dot{\sigma}}_{mP}$ versus $q$, respectively. Symbols and dashed lines represent the exact and phenomenological solutions, respectively. Parameters: $\Gamma_\nu=1$, $\beta_1=2$ and $\beta_2=1$.}
    \label{maximum_PowerAndSS}
\end{figure}

Although the behavior of efficiency is independent on $q$ (provided the system
is constrained in the ordered phase), it is worth highligthing
two important aspects about the system performance. First, 
the phase transition will also be marked by a discontinuity of $\eta$, whereas 
the absence of a phase transition in the heat-engine regime will imply
by the system achieving ideal efficiencies as $\langle {\cal P} \rangle=0$  and 
$V_1\beta_1=V_2\beta_2$
(see e.g. Fig.~\ref{Power_Grid} insets).
In addition,  the maximum power $-\mom{\mathcal{P}}_{mP}$ (with corresponding
dissipation $\mom{\dot{\sigma}}_{mP}$),  exhibits a  linear increase on  $q$,
as illustrated in Fig.~\ref{maximum_PowerAndSS}.

In a similar fashion, the performance of
BEG system can be characterized via following expressions 
\begin{align}
    \mom{\mathcal{P}}&=
        (V_2-V_1)\mathcal{J}_{1}(m,\rho)+(\Delta_2-\Delta_1)\mathcal{J}_{2}(m,\rho),\\
    \label{Power_General_Both}
    \mom{\dot{\sigma}}&=
        (\beta_2V_2-\beta_1V_1)\mathcal{J}_{1}(m,\rho)+(\beta_2\Delta_2-\beta_1\Delta_1)\mathcal{J}_{2}(m,\rho),
\end{align}
where $\mathcal{J}_{1}(m,\rho)$ and $\mathcal{J}_{2}(m,\rho)$ are corresponding fluxes evaluated in the NESS. Although exact, there is no closed form for them, since they depend in $m,\rho$ whose evaluations evolve  transcendental equations given by Eq.~(\ref{BEG_minimum}). However, we can consider an approximate description in
the regime $m\approx 1$ and $\rho\approx 1$, in which 
closed expressions for above fluxes are obtained, as shown in Appendix~\ref{apd} (a similar description can be done
for the Potts simply by setting $m \approx  1$ in Eqs.~(\ref{heat2})-(\ref{power})).
Fig.~\ref{Power_Grid} also illustrates, for the same parameters as the Potts case, the different regime operations for the BEG for $\Delta_1=\Delta_2=0$. Note that the engine regime range 
is the same in both cases (e.g.  independent on the model details), in consistency with Ref.~\cite{forao2025characterization}. However, there are two differences. The former is
(except for $q=2$),   Potts like interactions outperform those based on BEG ones, above all for $q=3$, provided the system is constrained in the ordered phase. Also, the heat-engine range in the former case ($V_2=3$)  is larger for the BEG
than the Potts one, because the system is constrained in the ordered phase in the former case but it is marked by a phase transition in the latter one.

\subsection{Different couplings $\Gamma_1\neq \Gamma_2$ and optimal  coupling ensuring maximal power}\label{different}
\begin{figure}
    \centering
    \includegraphics[width=1.\linewidth]{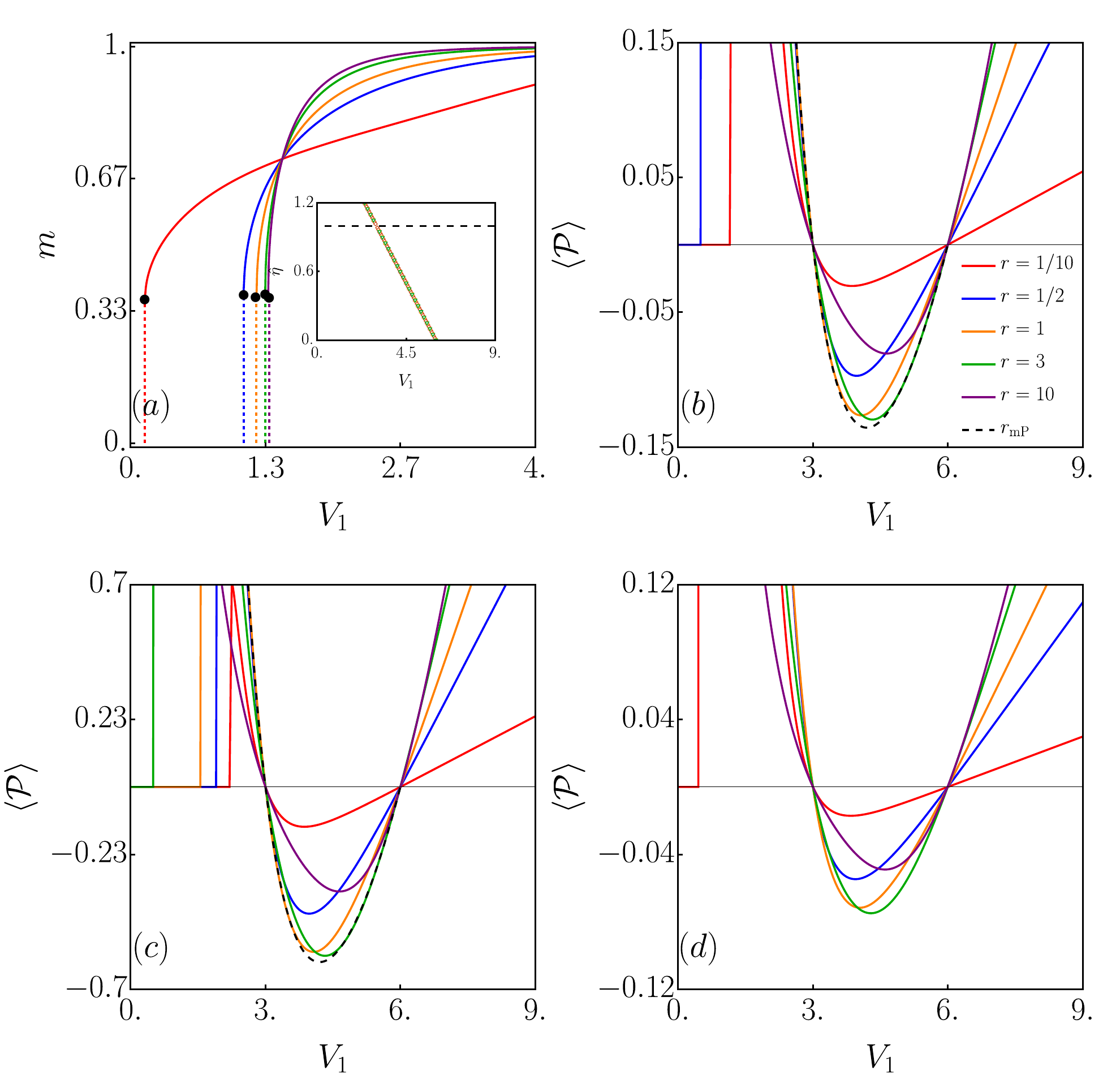}
    \caption{The influence of different couplings between thermal reservoirs. The depiction of the order-parameter $m$ (a) and power $\mom{\mathcal{P}}$ for the $q=3$ Potts model for different ratios $r\equiv(\Gamma+\Delta\Gamma)/(\Gamma-\Delta\Gamma)$, respectively. $r_{mP}$
    accounts to the ratio which maximizes $-\langle {\cal P}\rangle $. In  (c) and (d), the plot of $\mom{\mathcal{P}}$ for $q=10$ and BEG model 
    $\Delta_1=\Delta_2=0$, respectively. Parameters: $\Gamma=1$, $\beta_1=2$, $\beta_2=1$ and $V_2=6$.}
    \label{GridPowerMag_Potts_Asym}
\end{figure}

We now turn to the  $\Gamma_1\neq \Gamma_2$ case, meaning  different
coupling strengths between the system and thermal bath.
It also can be related with   activation energies of type $\Gamma_\nu=\Gamma e^{-{\beta_\nu}E^{(\nu)}_a/2}$  included
on the dynamics.
Although less explored than equal couplings, different couplings
can lead to remarkable different behavior of  quantities, such 
as different behaviors of power and dissipation \cite{proesmans2016linear,fiorek} or even a substantial influence on the
phase transition properties of collective engines \cite{gatien,mamede2023}. Starting with the Potts case and by performing likewise to Eq.~(\ref{ppp}), the NESS is characterized by the following
order-parameter expression 
\begin{widetext}
    \begin{equation}
    m=\frac{2 \left[\Gamma _1 \sinh \left(\frac{\beta_1V_1}{2}m\right)+\Gamma _2 \sinh \left(\frac{\beta_2V_2}{2}m\right)\right]}{\Gamma _1 \left[q \cosh \left(\frac{\beta_1V_1}{2}m\right)-(q-2) \sinh
   \left(\frac{\beta_1V_1}{2}m\right)\right]+\Gamma _2 \left[q
   \cosh \left(\frac{\beta_2V_2}{2}m\right)-(q-2) \sinh \left(\frac{\beta_2V_2}{2}m\right)\right]},
   \label{general_m_expansion_Coupling}
\end{equation}
\end{widetext}
irrespectively the temperatures of thermal baths, energy parameters, the value of $q$
and couplings strengths $\Gamma_\nu$'s. We pause again to make a few comments about Eq.~(\ref{general_m_expansion_Coupling}). First, 
it reduces to Eq.~(\ref{ppp}) for $\Gamma_1=\Gamma_2$. Second,
for $\Gamma_2\ll\Gamma_1$ and $\Gamma_1\ll\Gamma_2$, above expressions become simpler and approximately given by
\begin{align}
    m&=f_1(m)-r\left[\frac{2 q e^{\beta _1 m V_1} \sinh \left(\frac{1}{2} m \left(\beta _1 V_1-\beta _2 V_2\right)\right)}{\left(e^{\beta _1 m V_1}+q-1\right){}^2}\right],\\
    m&=f_2(m)+\frac{1}{r}\left[\frac{2 q e^{\beta _2 m V_2} \sinh \left(\frac{1}{2} m \left(\beta _1 V_1-\beta _2 V_2\right)\right)}{ \left(e^{\beta _2 m V_2}+q-1\right){}^2}\right],
\end{align}
respectively, where $r\equiv\Gamma_2/\Gamma_1$ and $f_\nu(m)\equiv 1-{q}/[{e^{m \beta _{\nu } V_{\nu }}+q-1]}$. Third,
expressions for the power and the entropy production are given by 
\begin{align}
\label{powc}
    \mom{\mathcal{P}}&=\frac{\Gamma _1 \Gamma _2  (q-1) \left(V_1-V_2\right)m e^{\frac{1}{2} \beta _2 m V_2} \left[e^{m \left(\beta _1 V_1-\beta _2 V_2\right)}-1\right]}{\Gamma _1
   \left(e^{\beta _1 m V_1}+q-1\right)+\Gamma _2 e^{\frac{1}{2} m \left(\beta _1 V_1-\beta _2 V_2\right)} \left(e^{\beta _2 m V_2}+q-1\right)},\\
   \mom{\dot{\sigma}}&=\frac{\Gamma _1 \Gamma _2  (q-1) \left(\beta_1V_1-\beta_2V_2\right)m e^{\frac{1}{2} \beta _2 m V_2} \left[e^{m \left(\beta _1 V_1-\beta _2 V_2\right)}-1\right]}{\Gamma _1
   \left(e^{\beta _1 m V_1}+q-1\right)+\Gamma _2 e^{\frac{1}{2} m \left(\beta _1 V_1-\beta _2 V_2\right)} \left(e^{\beta _2 m V_2}+q-1\right)},
\end{align}
respectively, once again delimiting the heat-engine regime between $V_1=V_2$ and $\beta_1V_1=\beta_2V_2$, such latter implying that $\langle {\dot \sigma}\rangle=0$ and $\eta=\eta_c$. Fourth,   the system efficiency $\eta$ is also
similar to the equal couplings case and exhibits the same linear dependence on the ratio $V_1/V_2$.
From Eqs.~(\ref{general_m_expansion_Coupling}) and (\ref{powc}), it is immediate to see that
different couplings between thermal
baths influence the phase transition properties and the
system performance, as depicted in Fig.~\ref{GridPowerMag_Potts_Asym}.  They move for lower $V_1$'s as the ratio $r$ is lowered (the other way around as
$V_2$ is varied). However they remain continuous and discontinuous for $q=2$ 
and $q\ge 3$, respectively. In the former case,  critical lines follow
the relation  $\Gamma_1\beta_1V_1+\Gamma_2\beta_2V_2=2(\Gamma_1+\Gamma_2)$, 
consistent with Eq.~(\ref{Phase_General1}) for $A_\nu=\Gamma_\nu\beta_\nu$
and $\Psi=2(\Gamma_1+\Gamma_2)$.
We also remark the existence of an optimal $r$ ensuring
superior $\langle{\cal P}\rangle$'s. To see this, we introduce the following parametrization $\Gamma_\nu=\Gamma+(-1)^\nu\Delta\Gamma$ and one resorts (for simplicity) to the phenomenological description (in a similar fashion to the BEG). By maximizing the power with respect to  coupling difference  $\Delta \Gamma$ we find that, for the parameter choice $\varphi\equiv\{q,V_1,V_2,\beta_1,\beta_2\}$, the optimal ratio $r_{\text{mP}}$ is given by

\begin{equation}r_{\text{mP}}=\frac{\Gamma+\Delta\Gamma(\varphi)_{\text{mP}}}{\Gamma-\Delta\Gamma(\varphi)_{\text{mP}}},
\end{equation}
where
$\Delta\Gamma(\varphi)_{mP}$ reads
\begin{equation}
    \Delta\Gamma(\varphi)_{mP}=\Gamma\frac{\left(\phi _1+\phi _2\right) \left[(q-1) \phi _1 \phi _2+1\right]}{\left(\phi _1-\phi _2\right) \left[(q-1) \phi _1 \phi
   _2-1\right]}+\mathfrak{h}\left(\varphi\right),
\end{equation}
with $\mathfrak{h}(\varphi)$ given by
\begin{equation}
    \mathfrak{h}(\varphi)=2 \Gamma\sqrt{\frac{\phi _1 \phi _2 \left[(q-1) \phi _1^2+1\right] \left[(q-1) \phi _2^2+1\right]}{ \left(\phi _1-\phi
   _2\right){}^2 \left[(q-1)\phi _1 \phi _2-1\right]{}^2}},
\end{equation}
and $\phi_\nu\equiv e^{-\beta_\nu V_\nu/2}$.  The existence of coupling ratio $r_{\text{mP}}$ depends on model parameters and on $q$. 
Lastly, we discuss some results for different couplings for the BEG.
Since expressions become substantially more involved in such a case, we shall curb ourselves
to the characterization of the criticality from a linear analysis
of Eq.~(\ref{mee}) given by $\overline{{\dot n}}_i = \sum_j B_{ij}\overline{n}_j$, with
$\{i,j\}\in \{\pm\}$, where $B$ denotes the Jacobian matrix. It has a  2x2 form
where elements
$B_{++}=-B_{-+}$ and $B_{+-}=B_{--}$ (not shown). Their eigenvalues  are given by $\lambda=\left(-B_{++}+B_{+-}\pm \sqrt{B^2_{++}+6B_{++}B_{+-}+B^2_{+-}}\right)/2$. Since  $B_{+-}$ is always negative, the criticality is obtained as $B_{++}=0$, leading to the Eq.~(\ref{Phase_General1}), where
parameters $A_1,A_2$ and $\Psi$ are given by
\begin{align}
A_1 &= 2\Gamma_1^2 S_1 x_2 + \Gamma_1\Gamma_2(x_1^2 +4 x_1 + x_2^2), \\
A_2 &= 2\Gamma_2^2 S_2 x_1 + \Gamma_1\Gamma_2(x_1^2 + 4x_2 + x_2^2), \\
\Psi &= (\Gamma_1 S_1 + \Gamma_2 S_2)[\Gamma_1 x_2 (x_1^2 + 2) + \Gamma_2 x_1 (x_2^2 + 2)],
\end{align}
where $x_\nu\equiv e^{\beta_\nu\Delta_\nu/2}$ and $S_\nu\equiv x_\nu+2$.
Note that $A_\nu=6(\Gamma_\nu^2+\Gamma_1\Gamma_2)$ and
$\Psi=9(\Gamma_1+\Gamma_2)^2$ as $\Delta_1=\Delta_2=0$.
The system performance for different couplings in also depicted in Fig.\ref{GridPowerMag_Potts_Asym}d, also showing the existence
of an optimal coupling. 
\subsection{Beyond the all-to-all case. Results for square-lattice topologies}\label{Simulations}

\begin{figure}[h!]
    \centering
    \includegraphics[width=1.\linewidth]{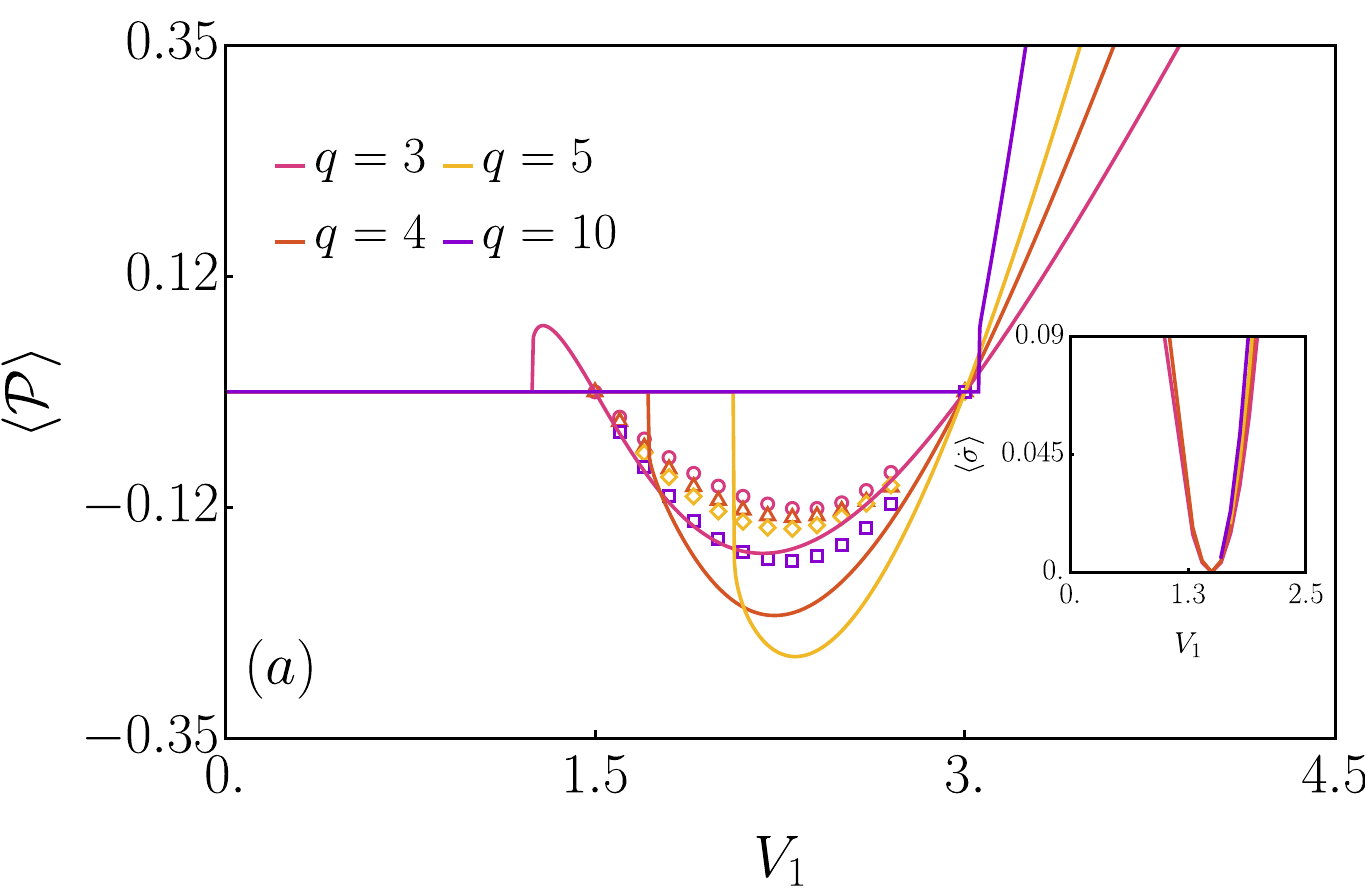}
     \includegraphics[width=1.\linewidth]{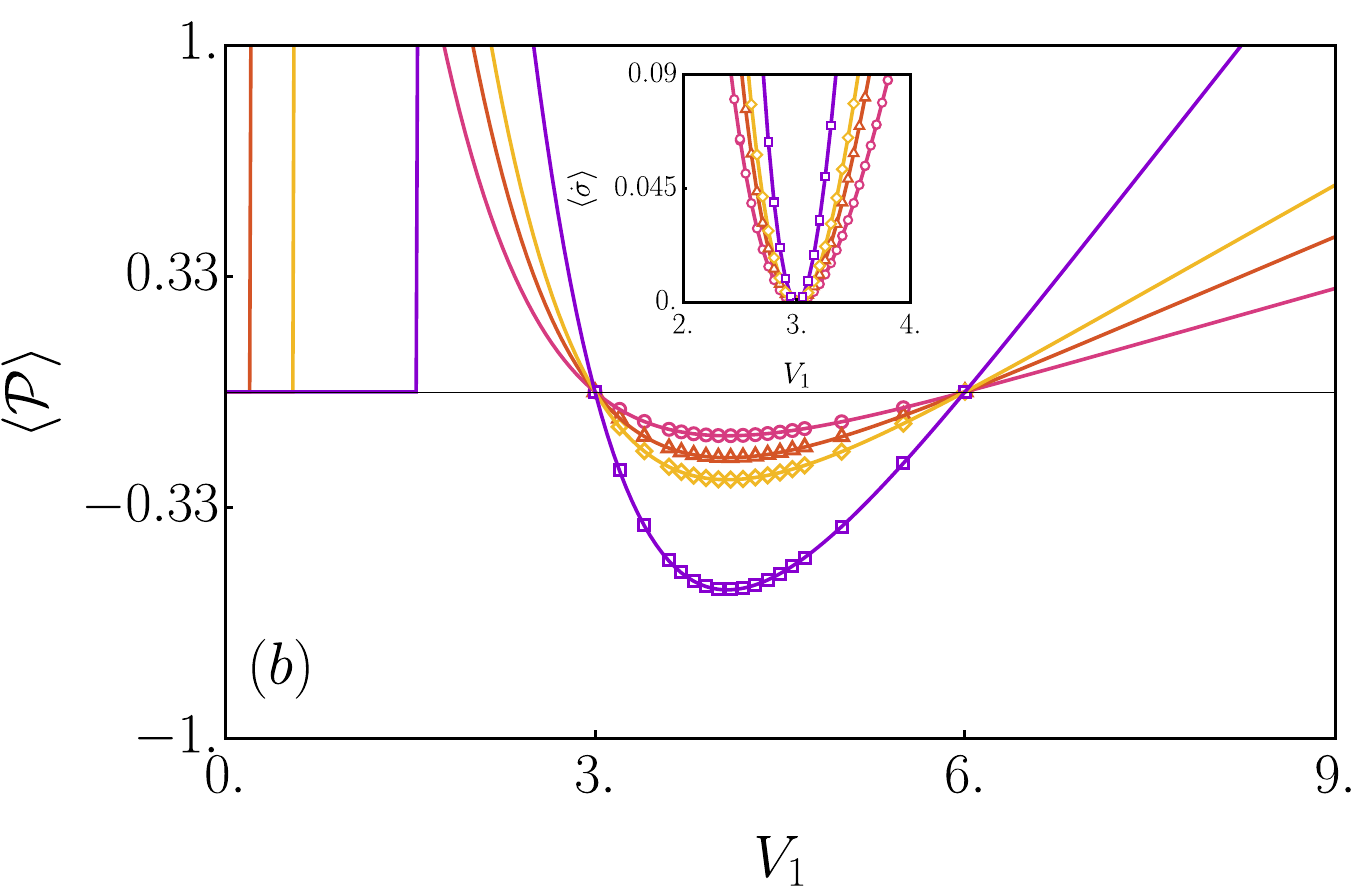}
    \caption{Depiction of the power $\mom{\mathcal{P}}$ (main panels) and entropy production (insets) for the Potts model in the square lattice (symbols) and the all-to-all (solid lines) for different values of $q$. Top and bottom panels we use $V_2=3$ and $V_2=6$, respectively. Parameters: $\Gamma_\nu=1$, $\beta_1=2$ and $\beta_2=1$.}
    \label{Square_Lattice_Power}
\end{figure}

In this section we advance over the all-to-all by investigating the system behaviors and engine peformances
for square-lattice topologies ($k=4$). Due the absence of exact results, we employ numerical simulations via Gillespie algorithm and phase transitions are characterized by by resorting to the finite size scaling. For simplicity, we shall curb ourselves for $\Gamma_1=\Gamma_2$. 
Fig.~\ref{Square_Lattice_Power} depicts the  the  $\langle{\cal P}\rangle$ and $\langle {\dot \sigma} \rangle$ for different $q$ together the comparison with the all-to-all case (solid lines). Results are very close to each other for the range of parameters in which the system is deeply constrained in the ordered phase ($m\approx 1$), showing that the role of topology is not important in such cases (differences between results are almost imperceptible), as exemplified for $V_2=6$. Conversely, they deviate  for $V_2=3$  because all quantities exhibit more dependence on the parameters and on the system size for $k=4$ and by the fact that the phase transitions occur "within" the engine regime for all-to-all interactions but not for $k=4$. Moreover, all-to-all interactions provide superior power-outputs and are more dissipative (not shown). As a last comment, it is also worth pointing out that efficiency also follows Eq.~(\ref{eff}), being also independent on the temperature and the lattice topology.

\section{Beyond the simultaneous contact between thermal reservoirs: A two box like approach}\label{nonsimult}

\begin{figure}
    \centering
    \includegraphics[width=1.\linewidth]{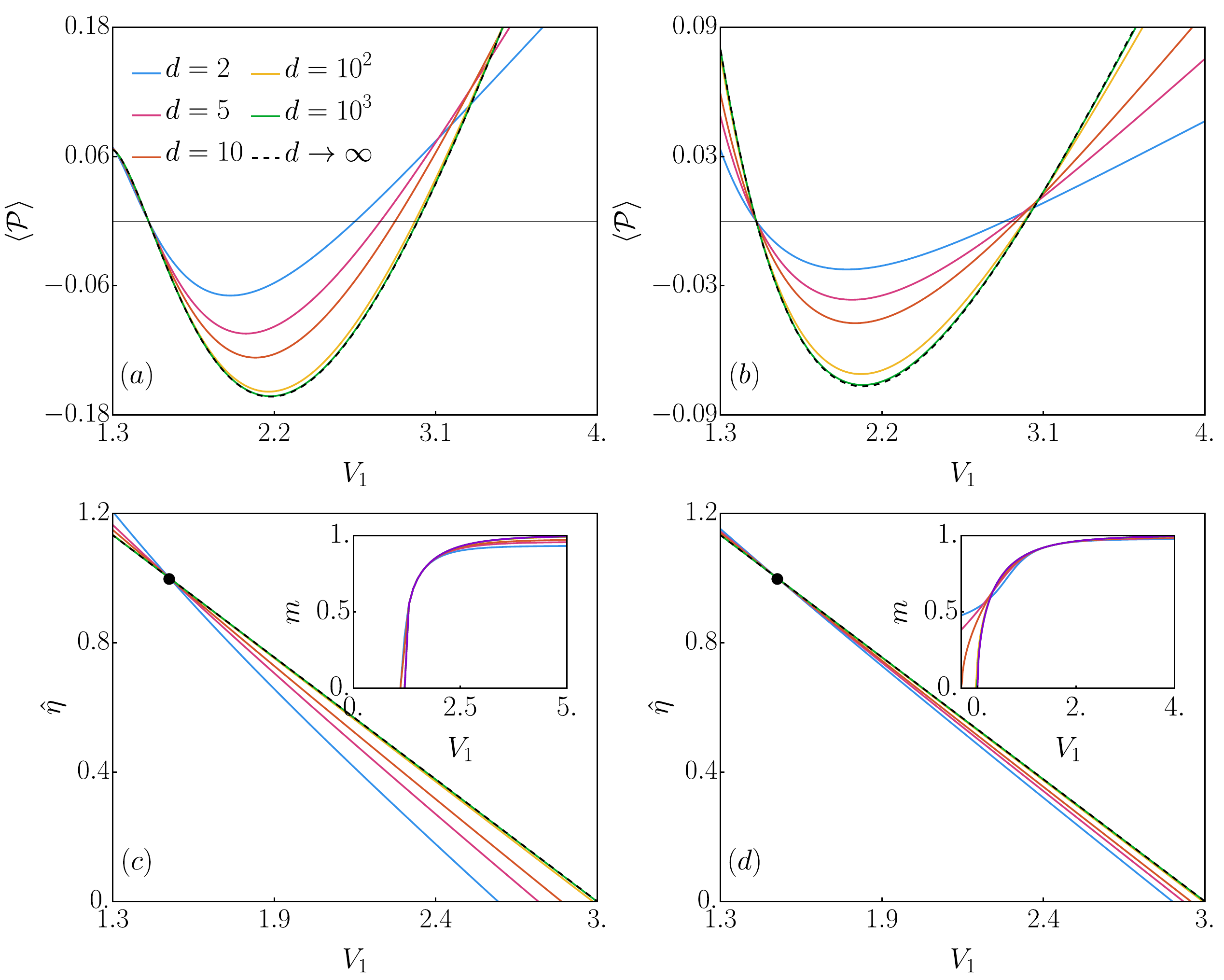}
    \caption{For the same parameters of Fig.~\ref{Figure_Phase_Transition}, top panels depict of the power $\mom{\mathcal{P}}$ versus $V_1$ for different $d$'s for the $q=3$ (left) and BEG (right). Bottom panels show the corresponding efficiencies $\eta$'s and $m$ (inset). Symbol $\bullet$ shows the crossover from heat-engine to pump regime in which $\eta=\eta_c$. Dashed lines denote the simultaneous contact with both thermal baths. Parameters: $\Gamma_1=\Gamma_2=1$, $\beta_1=2$, $\beta_2=1$ and $V_2=3$ and $\Delta_1=\Delta_2=0$ (right).}
    \label{last}
\end{figure}

Previous  analysis focused on each unit simultaneously placed in contact with both thermal baths. 
A key question that naturally arises is how this assumption influences the system behavior and overall the system performance.
We advance  beyond the simultaneous case, in which each unit is now placed in contact with a single thermal bath per time  but there is 
a switching rate $d$ between thermal baths. Among the different
approach for dealing with a non simultaneous contact, we take the two-box description,  \cite{danielPhysRevResearch.2.043257,liang2021dissipation,busiello2021dissipation} investigated in the context of nonequilibrium bipartite dynamics. By curbing ourselves to the all-to-all interactions  as $N\rightarrow \infty$,
the system is now characterized by the two-index variable $(\alpha,\nu)$, the former and latter describing the system state and the thermal bath, respectively, whose time evolution of  probability  ${p}^{(\nu)}_\alpha(t)$ reads
\begin{equation}\label{seq}
\dot{p}^{(\nu)}_\alpha(t)=\sum_{\alpha'\neq \alpha}J^{(\nu)}_{\alpha \alpha'}(t)+\sum_{\nu'\neq \nu}\mathcal{K}^{(\alpha)}_{\nu \nu'}(t),\end{equation} where the former  term in the right
is akin to the right side of Eq.~(\ref{mee}) and accounts to "inter-state" dynamics, in which $J^{(\nu)}_{\alpha \alpha'}(t)$ is  given by
\begin{equation} 
J^{(\nu)}_{\alpha \alpha'}(t)=   \omega^{(\nu)}_{\alpha \alpha'} {\bar n}^{(\nu)}_{\alpha'}(t) - \omega^{(\nu)}_{\alpha' \alpha} {\bar n}^{(\nu)}_\alpha(t), 
\label{mee2}
\end{equation}  
whereas the latter is given by
\begin{equation}
\mathcal{K}^{(\alpha)}_{\nu \nu'}(t)=\Omega_{\nu \nu'}  {\bar n}^{(\nu')}_{\alpha}(t)-\Omega_{\nu'\nu}  {\bar n}^{(\nu)}_\alpha(t)
\end{equation}
and accounts to the exchanges between thermal baths  at the same state $\alpha$. The expression for the entropy production $  \mom{\dot{\sigma}(t)}$ is then given by
\begin{equation}
    \mom{\dot{\sigma}(t)}=\sum_{\nu}\sum_{\alpha'>\alpha}J_{\alpha'\alpha}^{(\nu)}(t)X^{(\nu)}_{\alpha'\alpha}(t)+\sum_{\alpha}\mathcal{K}_{21}^{(\alpha)}(t)Y^{(\alpha)}_{21}(t),
    \label{eps_Current}
\end{equation}
where  $X^{(\nu)}_{\alpha'\alpha}$ and $Y_{21}^{(\alpha)}(t)$ denote the affinities given by
$X^{(\nu)}_{\alpha'\alpha}(t)=\log[\omega^{(\nu)}_{\alpha'\alpha}\overline{n}^{(\nu)}_{\alpha}(t)/\omega^{(\nu)}_{\alpha\alpha'}\overline{n}^{(\nu)}_{\alpha'}(t)]$ and $Y_{21}^{(\alpha)}(t)=\log[\Omega^{(\alpha)}_{21}\overline{n}^{(1)}_{\alpha}(t)/\Omega^{(\nu)}_{12}\overline{n}^{(2)}_{\alpha}(t)]$, respecrively. By considering symmetric switchings, $\Omega_{12}=\Omega_{21}=d$ for any $\alpha \in \{0,...,q-1\}$ (Potts) and $\alpha \in \{0,\pm\}$ (BEG), it is immediate to see that extreme cases $d\rightarrow 0$ and $d\rightarrow \infty$ (see e.g. Fig.~\ref{last}d) reduce to the equilibrium regime (for a single set of temperature and energies)  and the simultaneous  contact between both  thermal baths (by replacing
 $V_\nu \rightarrow 2V_\nu$ 
into transition rates $\omega^{(\nu)}_{\alpha \alpha'}$'s), respectively. The NESS is characterized by the time-independent
probabilities $\{{\bar n}^{(\rm st~ \nu)}_{\alpha'}\}$, in which Eq.~(\ref{eps_Current}) acquires the form
$\langle {\dot \sigma} \rangle=-\sum_\nu\beta_\nu\langle {\dot Q}_{\nu} \rangle$, where
$\mom{\dot{Q}_\nu}=\sum_{\alpha,\alpha'}\Delta\mathcal{E}^{(\nu)}_{\alpha'\alpha}\mathcal{J}^{(\nu)}_{\alpha',\alpha}$ and $\mathcal{J}^{(\nu)}_{\alpha',\alpha}$ are given by Eq.~(\ref{mee2}) evaluated in the NESS, in which
steady state probabilities satisfy that the condition $\sum_{\alpha}\overline{n}^{(\nu)}_{\alpha} = 1/2$. Under this, the order parameters $m_\nu$  associated to the contact with the $\nu$-th thermal reservoir for the Potts models can defined by extending previous relations for simultaneous contact case in the following form
\begin{equation}
\overline{n}^{(\nu)}_0=\frac{1}{q}\left[\frac{1}{2}+(q-1)m_\nu\right],~~~ \overline{n}^{(\nu)}_{\alpha\neq 0}=\frac{1}{q}{\left(\frac{1}{2}-m_\nu\right)}.
\end{equation}
By inserting them into Eq.~(\ref{seq}), the  steady-state solutions are given by
\begin{widetext}
\label{mag11}
    \begin{align}
    m_1&=\frac{\mathcal{A}_1 \Gamma _1 \left[d e^{\beta _2 m_2 V_2}+\Gamma _2 \left(\mathcal{A}_2+q\right)\right]+\mathcal{A}_2 \Gamma _2 d e^{\beta _1 m_1 V_1}}{2 \Gamma _2 d
   \left(\mathcal{A}_2+q\right) e^{\beta _1 m_1 V_1}+2 \Gamma _1 \left(\mathcal{A}_1+q\right) \left[d e^{\beta _2 m_2 V_2}+\Gamma _2
   \left(\mathcal{A}_2+q\right)\right]},\label{mag_1_sublattice}\\
   m_2&=\frac{\mathcal{A}_2 \Gamma _2 \left[d e^{\beta _1 m_1 V_1}+\Gamma _1 \left(\mathcal{A}_1+q\right)\right]+\mathcal{A}_1 \Gamma _1 d e^{\beta _2 m_2 V_2}}{2 \Gamma _2 d
   \left(\mathcal{A}_2+q\right) e^{\beta _1 m_1 V_1}+2 \Gamma _1 \left(\mathcal{A}_1+q\right) \left[d e^{\beta _2 m_2 V_2}+\Gamma _2
   \left(\mathcal{A}_2+q\right)\right]},\label{mag_2_sublattice}
    \end{align}
\end{widetext}
where $\mathcal{A}_\nu\equiv e^{2\beta_\nu  m_\nu V_\nu}-1$ and the system order-parameter is then given by  $m=m_1+m_2$. It is immediate to ser that the limit of $d\to\infty$ with $m_1=m_2=m/2$ and one recovers Eq.~\eqref{general_m_expansion_Coupling}. Expressions for the power and entropy production are then given by 
\begin{widetext}
    \begin{align}
    \mom{\mathcal{P}}&=\frac{\left(m_1 V_1-m_2 V_2\right)\left(\mathcal{A}_1-\mathcal{A}_2\right) \Gamma _1 \Gamma _2 d (q-1) }{\Gamma _2 d \left(\mathcal{A}_2+q\right) e^{\beta _1 m_1
   V_1}+\Gamma _1 \left(\mathcal{A}_1+q\right) \left[d e^{\beta _2 m_2 V_2}+\Gamma _2 \left(\mathcal{A}_2+q\right)\right]},\\
   \mom{\dot{\sigma}}&=\frac{\left(\beta _1 m_1 V_1-\beta _2 m_2 V_2\right)\left(\mathcal{A}_1-\mathcal{A}_2\right) \Gamma _1 \Gamma _2 d (q-1) }{\Gamma _2 d \left(\mathcal{A}_2+q\right)
   e^{\beta _1 m_1 V_1}+\Gamma _1 \left(\mathcal{A}_1+q\right) \left[d e^{\beta _2 m_2 V_2}+\Gamma _2 \left(\mathcal{A}_2+q\right)\right]},
\end{align}
\end{widetext}
respectively. Since 
above expressions are quite long, we shall characterize the phase transition only for $q=2$ in which they  become simpler. By performing a order-parameter series  series expansion in the NESS, we have that 
\begin{equation}
0 = \phi_1 m_1 +  \phi_3 m_1^3 + \dots,
\end{equation}
where the coefficients are obtained by expanding the  right side of Eq.~(\ref{mag_2_sublattice})
at $m_1 = m_2 = 0$.
The critical point is identified by the condition \( \phi_1 = 0 \), leading
to the following relation
\begin{equation}
        \Gamma_1\beta_1V_1+\Gamma_2\beta_2V_2=\frac{\Gamma_1\Gamma_2}{d}\left(2-\beta _1 V_1\right) \left(2-\beta _2 V_2\right)+2\left(\Gamma _1+\Gamma _2\right),
\end{equation}
which deviates of the bilinear form given by Eq.~(\ref{Phase_General1}).
However, it is immediate to see that one recovers the expression for the critical line for the simultaneous case as $d\rightarrow \infty $.
Since the main expressions are also very cumbersome for the BEG, we shall focus on the case as $\Delta_1=\Delta_2=0$ and $\Gamma_1=\Gamma_2$ in which they become somewhat
simpler.
In such case, the criticality acquires a simpler form and given by
\begin{equation}
\beta_1 V_{1}+\beta_{2} V_{2}=3+\frac{9}{2d}-\frac{3}{d}(\beta_1V_1+\beta_2V_2)+\frac{2\beta_1\beta_2}{d}V_1V_2
\end{equation}
which also deviates of the bilinear form from Eq.~(\ref{Phase_General1})  but approaches it as $d\rightarrow \infty$.
Fig.~\ref{last} extends previous analysis for finite $d$ by
taking the same parameters of Fig.~\ref{Figure_Phase_Transition}.

We remark three important differences. First,  finite switching rates $d$   meaningfully influence on the phase transition and the system performance. They imply at lower $\langle {\cal P}\rangle$'s
for finite $d$'s  (Fig.~\ref{last}a-b) and become equivalent to the simultaneous contact for  fast switchings ($d\gg1$). Second, while the engine regime is delimited
by $m_1V_1=m_2V_2$ and $\beta_1m_1V_1=\beta_2m_2V_2$, both expressions  are different from the simultaneous case. 
Third,
the efficiency $\eta$ is given by $\eta=1-(m_1V_1)/(m_2V_2)$ and
although exhibits
a  dependence approximately linear on $V_1$ 
(see e.g. Fig.~\ref{last}c-d), $\eta$ is
 also  affected by $d$, since  both $m_1$ and $m_2$ are given by Eq.~(\ref{mag_1_sublattice})-(\ref{mag_2_sublattice}) and also depend on $d$. 
Finally, order-disorder phase transitions exists  for all $d$'s for the Potts system and remain discontinuous. On the other hand, they are suppressed for the BEG and    small $d$'s.

\section{Conclusions}\label{Conclusions}
We proposed different  thermodynamic descriptions of collective systems in which parameters (individual and interaction) energies on assume different values due to the contact to each thermal reservoir. Different ingredients, such as the system-reservoir coupling, the
kind of interactions (Ising $\times$ BEG $\times$ Potts), the topology of interactions and the fact the system-reservoir contact is simultaneous or not,  were carefully investigated. Transition lines are expressed by a linear combination of interaction energies times reciprocal  temperatures for the simultaneous thermal contact baths but deviating from it when the contact is not simultaneous.  The number os states display an important influence upon the system performance, leading to the superior power outputs, whereas
the efficiency is independent on it. Also, results indicate that the simultaneous contact between thermal baths provides superior performances than the not-simultaneous
case and an optimized power can be obtained  
by properly choosing the coupling between thermal baths.
We believe that our results indicate an alternative route for designing
desirable performance of collective heat engines and how they are influenced
by the phase transition.

\section*{Acknowledgments}\label{Aknow}
We acknowledge the financial support from Brazilian agencies CNPq and FAPESP under grants 2023/17704-2, 2024/08157-0, 2024/03763-0, 2022/15453-0.
\bibliography{refs}

\appendix

\setcounter{equation}{0}
\setcounter{figure}{0}
\setcounter{table}{0}
\setcounter{page}{1}
\setcounter{section}{0}
\makeatletter
\renewcommand{\theequation}{A\arabic{equation}}
\renewcommand{\thefigure}{A\arabic{figure}}
\renewcommand{\citenumfont}[1]{A#1}

\section{All-to-all transitions}\label{apa}
As stated in the main text, for both $q$-state Potts and BEG models,  transition
rates depend  on the difference between number of entities at some state. Starting with the Potts model, the general structure of this energy gap follows that
\begin{equation}
    \Delta \mathcal{E}_{\alpha,\alpha'}^{(\nu)}=-\frac{V_\nu}{N}\left(1+N_\alpha-N_{\alpha'}\right).
    \label{DiffEnergy_AlltoAll_Meso}
\end{equation}
On the other hand, for the BEG, where $\mathcal{S}=\{0,\pm\}$, the only possible transitions are those described below
\begin{eqnarray}
    \Delta\mathcal{E}^{(\nu)}_{+,-}&=&-\frac{2V_\nu}{N}(N_{+}-N_{-}+1),\nonumber\\~~~\Delta\mathcal{E}^{(\nu)}_{+,0}&=&-\frac{V_\nu}{N}(N_{+}-N_{-})+\Delta_\nu,\\~~~\Delta\mathcal{E}^{(\nu)}_{-,0}&=&-\frac{V_\nu}{N}(N_{+}-N_{-})-\Delta_\nu\nonumber.
\end{eqnarray}
For $N\rightarrow\infty$, the energy differences can be expressed
in terms of the density of states $\overline{n}_\alpha=\lim_{N\to\infty}n_\alpha/N$,
as shown  below
\begin{equation}
    \Delta \mathcal{E}_{\alpha,\alpha'}^{(\nu)}=-V_\nu\left({\overline n}_\alpha-{\overline n}_{\alpha'}\right),
    \label{DiffEnergy_AlltoAll}
\end{equation}
for the Potts and
\begin{eqnarray}
    \Delta\mathcal{E}^{(\nu)}_{+,-}&=&-2V_\nu({\overline n}_{+}-{\overline n}_{-}),\nonumber\\~~~\Delta\mathcal{E}^{(\nu)}_{+,0}&=&-V_\nu({\overline n_+}-{\overline n_{-}})+\Delta_\nu,\\~~~\Delta\mathcal{E}^{(\nu)}_{-,0}&=&-V_\nu({\overline n}_{+}-{\overline n}_{-})-\Delta_\nu\nonumber,
\end{eqnarray}
for the BEG, respectively.

\section{Phenomenological description for the BEG and Potts models}\label{apd}
The phenomenological description for both models is based on the fact that  $m\approx1$ (for BEG it also means $\rho\approx1$).  From Eq.~(\ref{Power_General_Both}) in the main text, one has that
\begin{widetext}
\begin{align*}
    \mathcal{J}_{1}(\varphi;1,1) &= \mathcal{A} \phi_2 \bigg[-2 \Gamma_1 (\phi_1^2 +1)(\phi_1^4 - \phi_2^4) \gamma_2^2 - \Gamma_2 (\phi_1^2 -1)(\phi_2^2 +1) \phi_1^2 \phi_2 - \Gamma_1 (\phi_1^2+1)(\phi_1^4 - \phi_2^2) \phi_2 - \Gamma_2 \phi_1^2 (\phi_1^2 -1) \phi_2^2\bigg] \gamma_1^3 \\
    &+ \gamma_2 \phi_1 \bigg[2 \Gamma_2 (\phi_2^2 +1) (\phi_1^4 - \phi_2^4) \gamma_2^2 + 2(\Gamma_1 + \Gamma_2) \phi_2 (\phi_1^4 - \phi_2^4) \gamma_2 + \Gamma_1 \phi_2^2 (\phi_1^4 - \phi_2^2)\bigg] \gamma_1^2 \\
    &+ \gamma_2^2 \phi_1^2 \bigg[\Gamma_2 \phi_2 (\phi_1^2 - \phi_2^4) + \gamma_2 \big(\Gamma_2 (\phi_2^2 +1)(\phi_1^2 - \phi_2^4)  - \Gamma_1 (\phi_1^2+1) \phi_2^2 (\phi_2^2 -1)\big)\bigg] \gamma_1 \\
    &- \gamma_2^3 \Gamma_1 \phi_1^3 \phi_2^2 (\phi_2^2 -1),
\end{align*}

\begin{align*}
    \mathcal{J}_{2}(\varphi;1,1) &= \mathcal{A}\phi_2 \bigg[2 \Gamma_1 (\phi_1^2 +1)(\phi_1^4 - \phi_2^4) \gamma_2^2 + \Gamma_2 (\phi_1^2 -1)(\phi_2^2 +1) \phi_1^2 \phi_2  + \Gamma_1 (\phi_1^2+1)(\phi_1^4 - \phi_2^2) \phi_2 + \Gamma_2 \phi_1^2 (\phi_1^2 -1) \phi_2^2\bigg] \gamma_1^3 \\
    &+ \gamma_2 \phi_1 \bigg[2 \Gamma_2 (\phi_2^2 +1) (\phi_2^4 - \phi_1^4) \gamma_2^2 + 2(\Gamma_1 + \Gamma_2) \phi_2 (\phi_2^4 - \phi_1^4) \gamma_2  + \Gamma_1 \phi_2^2 (\phi_2^2 - \phi_1^4)\bigg] \gamma_1^2 \\
    &+ \gamma_2^2 \phi_1^2 \bigg[\Gamma_2 \phi_2 (\phi_2^4 - \phi_1^2) + \gamma_2 \big(\Gamma_1 (\phi_1^2+1) (\phi_2^2 -1) \phi_2^2  + \Gamma_2 (\phi_2^2 +1)(\phi_2^4 - \phi_1^2)\big)\bigg] \gamma_1 \\
    &+ \gamma_2^3 \Gamma_1 \phi_1^3 \phi_2^2 (\phi_2^2 -1),
\end{align*}
where $\mathcal{A}=\Gamma_1\Gamma_2\gamma_1^2\gamma_2\phi_1\phi_2/D$, and
\begin{align*}
    D &= \gamma_2 \Gamma_1 \phi_2 \bigg[\Gamma_2 \phi_2 (\phi_1^2 \phi_2^2+1) \phi_1^2 + \gamma_2 (\phi_1^2+1) \big(\Gamma_2 (\phi_2^4+1) \phi_1^2+ \Gamma_1 (\phi_1^4+1) \phi_2^2\big)\bigg] \gamma_1^3 \\
    &+ \phi_1 \bigg[\Gamma_2 (\phi_2^2+1) \big(\Gamma_2 (\phi_2^4+1) \phi_1^2 + \Gamma_1 (\phi_1^4+1) \phi_2^2\big) \gamma_2^3 \\
    &\quad + \big(\Gamma_1^2 (\phi_1^4+1) \phi_2^3 + \Gamma_2^2 \phi_1^2 (\phi_2^4+1) \phi_2\big) \gamma_2^2 + \Gamma_2 \phi_2^2 \big(\Gamma_2 (\phi_2^2+1) \phi_1^2 + \Gamma_1 (\phi_1^4+\phi_2^2)\big) \gamma_2 + \Gamma_2^2 \phi_1^2 \phi_2^3\bigg] \gamma_1^2 \\
    &+ \gamma_2 \Gamma_1 \phi_1^2 \phi_2 \bigg[\Gamma_2 \phi_2 (\phi_1^2 \phi_2^2+1) \gamma_2^2 + \big(\Gamma_1 (\phi_1^2+1) \phi_2^2 + \Gamma_2 (\phi_2^4+\phi_1^2)\big) \gamma_2  + \Gamma_2 \phi_2 (\phi_1^2+\phi_2^2)\bigg] \gamma_1+ \gamma_2^2 \Gamma_1^2 \phi_1^3 \phi_2^3,
\end{align*}
\end{widetext}
where
$\phi_\nu\equiv\exp(-\beta_\nu V_\nu/2)$ and $\gamma_\nu\equiv\exp(-\beta_\nu\Delta_\nu/2)$. Despite cumbersome, they solely depend on the model
parameters. By inserting them on Eqs.~(\ref{Power_General_Both}), thermodynamics quantities
can be (approximately) evaluated.

\end{document}